\documentclass[useamsfonts]{pasj00}
\usepackage{bm}

\begin{document}
\SetRunningHead{Takeuchi et al.}{Leonid Meteor Storm: Self-Organizing 
State Space Approach}
\Received{2002 October 30}
\Accepted{2002 December 15}

\title{Application of a Self-Organizing State Space Model to the Leonid 
Meteor Storm in 2001}

\author{%
Tsutomu T. \textsc{Takeuchi}\thanks{Current address: Division of Particle and Astrophysical
Science, Nagoya University, Furo-cho, Chikusa-ku, Nagoya 464-8602, Japan}}

\affil{Optical and Infrared Astronomy Division, 
National Astronomical Observatory,\\
2--21--1, Osawa, Mitaka, Tokyo 181--8588}

\email{takeuchi.tsutomu@g.mbox.nagoya-u.ac.jp}

\author{Shigetomo \textsc{Shiki}}

\affil{Image Information Division, Advanced Computing Center,\\
The Institute of Physical and Chemical Research,\\
2-1 Hirosawa, Wako, Saitama 351--0198}

\email{shiki@riken.go.jp}

\author{Daisuke \textsc{Miyamoto}}

\affil{Shibuya Kyoiku Gakuen Senior \& Junior High School
1--3, Wakaba, Mihama, Chiba 261--0014}

\author{Hideaki \textsc{Fujiwara}}

\affil{College of Arts and Sciences, The University of Tokyo, 
3--8--1 Komaba, Meguro, Tokyo 153--8914}

\author{Jun \textsc{Kitazume}}

\affil{Sugamo Highschool, 1--21--1 Kamiikebukuro, Toshima, Tokyo 170--0012}

\and

\author{Yousuke \textsc{Utsumi}}
 
\affil{Secondary School Attached to Faculty of Education of 
the University of Tokyo,\\
1--15--1 Minami-dai, Nakano, Tokyo 164--8654}


\maketitle

\begin{abstract}
The Leonids show meteor storms in a period of 33 years, and known as 
one of the most active meteor showers.
It has recently shown a meteor stream consisting of several narrow dust trails
made by meteoroids ejected from a parent comet.
Hence, an analysis of the temporal behavior of the meteor flux is 
important to study the structure of the trails.
However, statistical inference for the count data is not an easy task,
because of its Poisson characteristics.
We carried out a wide-field video observation of the Leonid meteor 
storm in 2001.
We formulated a state-of-the-art statistical analysis, which is called 
a self-organizing state space model, to infer the true behavior of the dust 
density of the trails properly from the meteor count data.
{}From this analysis, we found that the trails have a fairly smooth spatial
structure, with small and dense clumps that cause a temporal burst of 
meteor flux.
We also proved that the time behavior (trend) of the fluxes of bright meteors
and that of faint meteors are significantly different.
In addition we comment on some other application of the self-organizing 
state-space model in fields related to astronomy and astrophysics.
\end{abstract}

\section{Introduction}

\subsection{Leonid Meteor Storm as a Stochastic Counting Process}

The Leonids show meteor storms in a period of 33 years, and known 
as one of the most active meteor showers.
It is believed that the Leonid meteor storm occurs when the Earth 
approaches the orbit of the dust trails of the comet 55P/Tempel--Tuttle. 
Many authors have predicted that the Leonid meteor storm would occur in 2001
(see e.g., \cite{watanabe02} and references therein).  
It has been recognized that a meteor stream consists of several narrow dust 
trails, each of which is made by meteoroids ejected at a particular return
of the parent comet (e.g., \cite{watanabe97,brown02}).
Hence, the temporal behavior of the meteor flux provides important 
information of the spatial and density structure of the dust trails.

We carried out a wide-field video observation of the Leonid meteor 
storm in 2001 \citep{shiki03}, and found the following results through 
a statistical analysis:
\begin{enumerate}
\item The time variation of the hourly rate (HR) shows many small peaks.
\item The HR profile of bright meteors and that of faint meteors are 
significantly different.
\item The small peaks are associated with a burst of faint meteor flux.
\item Some signatures of Poisson-like process are found in the interval 
distribution of meteors and in their autocorrelation function.
\end{enumerate}
Many previous studies tried to analyze the time variation of the meteor
flux (e.g., \cite{watanabe97} and references therein).
However, it is not a trivial task to infer the true behavior of the trail
density, since it suffers from a strong statistical fluctuation.
The difficulty is due to the fact that measuring the meteor flux is
a kind of typical counting process, whose fluctuation is basically 
not Gaussian, but is characterized by a Poisson distribution.
The Poisson counting process has a variance equal to its mean value,
i.e., when the mean is large, the associating variance is also large.
This character prevents us from applying the popular classical time series 
analysis methods to the count data, because some basic assumptions are 
utterly violated (e.g., \cite{brockwell96,cameron98}).

\subsection{Development of Time Series Analysis}

Classical time series analyses have often appeared in an astronomical 
context.
Data from astronomy as well as from other physical, biological, or econometric
studies consist of a sequence of numbers, $\{x_1, x_2, x_3, \cdots, x_T\}$,
obtained by measuring the quantity $x_t$ during a sequence of times.
The subscripts represent discrete time that runs from $t=1$ to $T$.
This discrete treatment of time is sufficient for a variety of practical 
applications.\footnote{
Here, we concentrate on a set of equally sampled time series data, but,
as mentioned later, the method presented in this article can also be 
applied to irregularly sampled data.
}
A comprehensive summary of the classical time series analysis and its
applications to astronomical datasets can be found in \citet{scargle81}.

Today, in order to handle a wider range of time series, including some 
latent variables, {\sl the state space model} has been proposed from 
the field of engineering and system optimal control.
Especially, the merit of using the state space form is that it can properly
treat time-varying parameters in the system.
We note that classical time series models can be defined as special cases 
of the general state space model.
The well-known Kalman filter \citep{kalman60} gives an algorithm to
estimate the system parameters recursively, i.e., it gives one-step-ahead
estimations (called `filtering') everytime we have a new data 
point in the series.

A Kalman filter assumes the Gaussianity for the system noise terms, which
is not suitable for the data we often encounter in astronomical 
applications.
Further, as seen in section~\ref{sec:state}, the Kalman filter is designed 
for a linear system.
A generalized state space model with nonlinearity and non-Gaussianity 
has been proposed to overcome these shortcomings (\cite{kg96}, and references
therein).
An excellent guide for their applications can be found in \citet{brockwell96}.

However, there still remains an additional difficulty concerning astronomical 
data analysis.
Astronomers frequently meet a situation in which they must handle 
a dataset with small counts, such as a very low-level signal of the faintest 
sources.
Count data introduce complications of discreteness and 
heteroskedasticity.\footnote{This technical term means that the variance is 
not constant.}
The inclusion of zero counts appears to be a pitfall to apply, e.g., usual 
regression methods.
Despite its frequency that we should tackle such datasets, 
time series models for counting data are in their infancy, yet remarkably 
many models have been developed.
\citet{cameron98} provides a through discussion on general counting problems, 
including time series counting data.

Now we close the chronicle of time series analysis.
The development is concisely summarized in \citet{kitagawa01}.
In this paper, we propose a suitable method of analyzing time series count 
data, which sometimes have zeros in the sequence.
This approach, the self-organizing state space model, has been 
developed only very recently (\cite{kitagawa98,higuchi99}, and references 
therein).
It makes extensive use of a large computing power of modern computers.

In this paper we statistically formulate the count data of the Leonid meteor 
storm.
The count rate obeys a Poisson process with a latent, time-varying 
Poisson intensity, $\lambda_t$.
We made an attempt to estimate the temporal behavior of the hidden parameter
$\lambda_t$ from the observed count data.
We should note that our data have some gaps in observations caused by the 
length of video tapes and other reasons.
Our method can easily overcome such gaps, and properly infer the value 
in the observational time gaps.

The rest of the present paper is as follows. 
In section~\ref{sec:method}, we formulate a self-organizing state space
model of the Leonid count data. 
We start with a linear state space model, and then develop toward more
general methods.
We present the results and discussions in section~\ref{sec:results}.
section~\ref{sec:conclusion} is devoted to a summary.

\section{Data}\label{sec:data}

The original data consist of the magnitudes and observed time of the meteors,
which were recorded on a video tape.
Although they include some sporadic meteors and meteors belonging to other 
meteor showers, here we concentrate only on the Leonids in the analysis.
Detailed data descriptions are found in \citet{shiki03}.

In the usual manner of the meteor-shower analysis, an hourly rate (HR) is used 
to describe the time variation of the meteor flux.
However, in this study, it is more appropriate to present the data based on
the count rate per minute; hence, we use the count rate [min$^{-1}$] 
throughout this paper.

\section{Method}\label{sec:method}

\subsection{State Space Model}\label{sec:state}

A classical state space model consists of the following equations:
\begin{eqnarray}
  x_t &=& {\sf F}_t x_{t-1} + {\sf G}_t v_t \, , \quad v_t \sim N(0,{\sf Q}_t)
   \, ,\label{eq:state}\\
  y_t &=& {\sf H}_t x_t + w_t \,, \quad w_t \sim N(0,{\sf R}_t)\, .
  \label{eq:obs}
\end{eqnarray}
Here, $x_t$ is called a state vector, which represents the (unobserved)
state of the system, and $y_t$ is the observed sequence of data.
The idea underlying the model is that the development of the system over 
time is determined by $x_t$ according to equation~(\ref{eq:state}).
However, because $x_t$ cannot be observed directly, we must base 
the analysis on observations, $y_t$.
We call equation~(\ref{eq:state}) the state equation, and 
equation~(\ref{eq:obs}) the observation equation.
The error terms, $v_t$ and $w_t$, are distributed according to Gaussian
probability distribution functions, $N(0,{\sf Q}_t)$ and $N(0,{\sf R}_t)$, 
respectively, where ${\sf Q}_t$ and ${\sf R}_t$ are covariance matrices.

Matrices ${\sf F}_t$, ${\sf G}_t$, ${\sf H}_t$, ${\sf Q}_t$, and ${\sf R}_t$ 
are initially assumed to be known, and the error terms, $v_t$ and $w_t$, are 
assumed to be serially independent and independent of each other at all time.
In practice, some or all of the matrices depend on the elements of an unknown 
parameter vector, $\theta$.

A considerable advantage of the state space approach is the ease with which
missing observations can be dealt with.
The estimation problems in time series analysis can be classified into the
following three categories with respect to the dependence on the observed 
data $\{y_1, \cdots, y_t \}$ to estimate the state vector $x_t$:
\begin{enumerate}
\item $y_{1:t-1}\equiv\{y_1, \cdots, y_{t-1}\}$ : prediction,
\item $y_{1:t}\equiv\{y_1, \cdots, y_t\}$ : filtering, and
\item $y_{1:u}\equiv\{y_1, \cdots, y_u\}\; (u > t)$ : smoothing.
\end{enumerate}

Kalman recursion equations give a one-step-ahead prediction of the state 
vector, $x_t$, and its error covariance matrix, by using 
$\{y_1, \cdots, y_{t-1} \}$, and filter the series when we have new data, 
$y_t$.
We do not go any further into the details of the Kalman filter here. 
For implementation, see, e.g., \citet{harvey81}, \citet{brockwell96} and 
\citet{durbin01}.
If we have a missing in data sequence, we simply perform prediction 
without filtering, and go to the next step.
This point is extensively discussed by \citet{akaike99}.

\subsection{Generalized State Space Model}

We can consider a nonlinear non-Gaussian state space model as being 
an extension of the linear case \citep{kitagawa87}:
\begin{eqnarray}
  x_t &=& F_t(x_{t-1},v_t) \; , \\
  y_t &=& H_t(x_t,w_t)\;.
\end{eqnarray}
Again, the first is the state equation and the second is the observation
equation.
These times, $v_t$ and $w_t$, are the system and observation noise with
non-Gaussian densities, $q_t(v_t)$ and $r_t(w_t)$, respectively.
The initial state $x_0$ is assumed to be distributed with the probability 
density $p_0(x_0)$.
Functions $F_t(x,v)$ and $H_t(x,w)$ are nonlinear ones of the state 
vector and noise \citep{brockwell96, durbin01, kitagawa01}.

It is convenient to express the model in a general form based on the
conditional distributions:
\begin{eqnarray}
  x_t &=& Q_t(\cdot | x_{t-1})  , \\
  y_t &=& R_t(\cdot | x_t)\;.
\end{eqnarray}
With this general state space model, we can handle discrete-valued time
series as well as discrete-state models.

For general state space models, the conditional distributions become 
non-Gaussian and their distributions cannot be completely specified
by the mean vectors and the covariance matrices, which is different
from the case of a Gaussian linear state space model and a Kalman filter.
A non-Gaussian filter is expressed as follows:
\begin{eqnarray}
  {\sf p}(x_t|y_{1:t-1})&=&\int {\sf p}(x_t|x_{t-1})
    {\sf p}(x_{t-1}|y_{1:t-1})dx_{t-1} \; , \\
  {\sf p}(x_t|y_{1:t}) &=& \frac{{\sf p}(y_t|x_t)
    {\sf p}(x_t|y_{1:t-1})}{{\sf p}(y_t|y_{1:t-1})} \; ,
\end{eqnarray}
where ${\sf p}(y_t|y_{1:t-1})$ is the predictive distribution of $y_t$,
\begin{eqnarray}
  {\sf p}(y_t|y_{1:t-1}) = \int {\sf p}(y_t|x_t){\sf p}(x_t|y_{1:t-1})dx_t \; .
\end{eqnarray}
However, a direct implementation of the formula requires computationally 
intense numerical integration that can only be feasible for some limited 
case.

\subsection{Monte Carlo Filter}

Instead of approximating the distribution and performing heavy numerical
integration, we can use Monte Carlo filtering, i.e., producing a large 
number of realizations which can be extracted from the distribution
\citep{kitagawa96}.
This method needs much less computational power.

The procedure is as follows:
\begin{enumerate}
\item Generate a random number, $x_0^{(j)} \sim p_0(x_0)$, for $j=1,\cdots,N$.
\item Repeat the following steps for $t=1,\cdots,T$:
  \begin{enumerate}
  \item Generate a random number $v_t^{(j)} \sim q(v)$ for $j=1,\cdots,N$.
  \item Compute $p_t^{(j)} = F[x_{t-1}^{(j)}, v_t^{(j)}]$ for 
    $j=1,\cdots,N$.
  \item Compute $w_t^{(j)} = {\sf p}[y_t|p_t^{(j)}]$ for $j=1,\cdots,N$.
  \item Generate $x_t^{(j)}$ ($j=1,\cdots,N$) by resampling of $p_t^{(j)}$
    ($j=1,\cdots,N$).
  \end{enumerate}
\end{enumerate}

In step 2(b), $p_t^{(j)} = F[x_{t-1}^{(j)}, v_t^{(j)}]$ can be 
considered to be independent realizations from the predictive distribution, 
${\sf p}(x_t|y_{1:t-1})$.
Given the observation $y_t$ and the realization $p_t^{(j)}$, we obtain
the importance weight, $w_t^{(j)}$ [step 2(c)], i.e., the likelihood
with respect to the observation, $y_t$.
The posterior probability of the realization is
\begin{eqnarray}
  &&{\sf Prob}[x_t = p_t^{(j)}|y_{1:t}] \nonumber \\
  &&= {\sf Prob}[x_t = p_t^{(j)}|y_{1:t-1},y_t] \nonumber \\
  &&= \frac{
    {\sf p}[y_t|p_t^{(j)}] {\sf Prob}[x_t=p_t^{(j)}|y_{1:t-1}]}{
    \sum_{k=1}^{N}
    {\sf p}[y_t|p_t^{(k)}] {\sf Prob}[x_t=p_t^{(k)}|y_{1:t-1}]}\nonumber\\
  &&= \frac{w_t^{(j)}}{\sum_{k=1}^{N}w_t^{(k)}} \;.
\end{eqnarray}
This means that the cumulative distribution function of 
${\sf Prob}[x_t=p_t^{(j)}|y_{1:t}]$ can be expressed by a step function,
\begin{eqnarray}
  {\sf Prob}[x_t=p_t^{(j)}|y_{1:t}]=\frac{1}{\sum_{k=1}^{N}w_t^{(k)}}
    \sum_{j=1}^{N} w_t^{(k)} I[x,p_t^{(j)}] \; ,
\end{eqnarray}
which has jumps at $p_t^{1}, \cdots, p_t^{N}$ with step sizes $w_t^{(1)}, 
\cdots, w_t^{(N)}$.
Here, $I(x,z)$ is the Heaviside function, which has a unit jump at $x=z$.

For the next step of the prediction, we must represent this distribution 
function by an empirical distribution with the form 
\begin{eqnarray}
  {\sf Prob}[x_t=p_t^{(j)}|y_{1:t}]=\frac{1}{N}
    \sum_{j=1}^{N} I[x,p_t^{(j)}] \; .
\end{eqnarray}
This can be done by resampling of $\{p_t^{(1)}, \cdots, p_t^{(N)}\}$ with
probabilities, 
\begin{eqnarray}
  {\sf Prob}[x_t=p_t^{(j)}|y_{1:t}]=
    \frac{w_t^{(j)}}{\sum_{k=1}^{N}w_t^{(k)}} \;.
\end{eqnarray}
We calculate this step using the von Neumann's acception--rejection method
(see e.g., \cite{knuth98} for details).

One of the main purposes of the time series analysis is to estimate
some hidden parameters, $\theta$, from the observed data.
The likelihood of the time series model specified by the parameter 
$\theta$ is obtained by 
\begin{eqnarray}
  L(\theta)={\sf p}(y_1, \cdots, y_T)=
    \prod_{t=1}^{T}{\sf p}(y_t|y_{1:t-1};\theta)\;.
\end{eqnarray}
Along with the above discussion, we can approximate the conditional density by 
\begin{eqnarray}
    {\sf p}(y_t|y_{1:t-1},\theta) \simeq \frac{1}{N} \sum_{j=1}^{N}w_t^{(j)}\;.
\end{eqnarray}

\subsection{Self-Organizing State Space Model}

In principle, the maximum likelihood estimate of $\theta$ is obtained by 
maximizing the log-likelihood, $\log L(\theta)$.
However, in practice the sampling error and long computational time 
often renders the direct maximum likelihood method impractical.
{}To remedy this problem, \citet{kitagawa98} proposed a sophisticated
method. 
Instead of estimating the parameter $\theta$ by the maximum likelihood, 
we consider a Bayesian estimation by augmenting the state vector, $x_t$, with
an unknown parameter, $\theta$, as
\begin{eqnarray}\label{eq:gsv}
  z_t = \left( 
    \begin{array}{@{\,}c@{\,}}
      x_t \\
      \theta_t
    \end{array}
  \right)\;, 
\end{eqnarray}
and construct the state space model for $z_t$ as
\begin{eqnarray}
  z_t = F^*(z_{t-1},v_t) \; , \\
  y_t \sim R(\cdot | z_t) \;.
\end{eqnarray}
This is called `the self-organizing state space model'.
In equation~(\ref{eq:gsv}), the parameter is unknown, but constant such that 
$\theta_t=\theta_{t-1} = \cdots = \theta$.
This can immediately extended to the time-varying (hyper)parameter case,
but we do not go any further here.

\subsection{Model for the Leonid Meteor Storm}

Here, we formulate our problem to estimate the temporal trend of 
the Leonid meteor shower hidden by a statistical fluctuation, which may be 
Poissonian.
For the Poisson count process, the observation equation is expressed as
\begin{eqnarray}\label{eq:poisson}
  y_t \sim \mbox{Poisson}(\lambda_t)=
    \frac{\lambda_t^{y_t} e^{-\lambda_t}}{y_t !} \; , \; t=1,\cdots,T \;,
\end{eqnarray}
\citep{higuchi99,higuchi01,kitagawa01}.
The system model becomes quite simple, as
\begin{equation}
  x_t = \left( 
    \begin{array}{@{\,}c@{\,}}
      \mu_t \\
      \log \sigma_{\mu,t}^2
    \end{array}
  \right) = \left( 
    \begin{array}{@{\,}c@{\,}}
      \mu_{t-1} \\
      \log \sigma_{\mu,t-1}^2
    \end{array}
  \right) + \left( 
    \begin{array}{@{\,}c@{\,}}
      v_t \\
      0
    \end{array}
  \right) \;,
\end{equation}
where $\mu_t\equiv\log \lambda_t$.
We adopt a first-order smooth trend such that $\mu_t = \mu_{t-1} + v_t$, 
$v_t \sim N(0,\sigma_\mu^2)$.
This treatment enables us to handle an arbitrary trend, because it
only assumes that the first-order difference of the trend is small.
The parameter $\sigma_\mu^2$ ($\log \sigma_\mu^2$) is simultaneously 
estimated by a recursive procedure.\footnote{
The logarithm is defined as $\log x \equiv \log_{10} x$.
It is set to conserve the positive definiteness of the trend.}
As recommended by \citet{higuchi01}, we set $\log \sigma^2_\mu \sim U([-6.0,-2.0])$
as the initial distribution of $\sigma^2_\mu$, where $U([a,b])$ denotes the uniform distribution
between $a$ and $b$.

As already mentioned above, Leonid meteors seem to behave like 
a Poisson process locally, and hence this model is appropriate to 
describe this data (see \cite{shiki03}).

\section{Results and Discussions}\label{sec:results}

The observations were made from $14^{\rm h}41^{\rm m}$ to 
$20^{\rm h}03^{\rm m}$~UT (322 minutes), hence $T=322$ in 
equation~(\ref{eq:poisson}).

\subsection{Global Behavior}

\begin{figure}
  \begin{center}
  \FigureFile(80mm,80mm){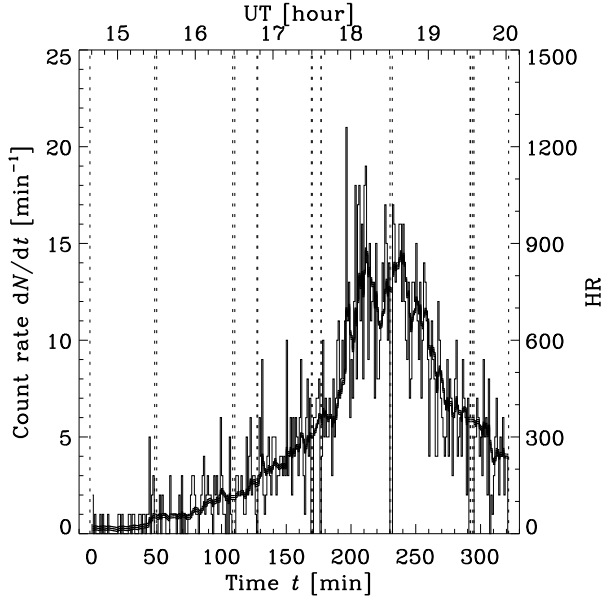}
  \end{center}
  \caption{Count rate data of the Leonid meteor storm and its estimate for
  the true density distribution.
  The vertical dotted lines indicate the boundaries of the data.
  Data are missing in small intervals between the dotted lines.
  The observed counts are shown by the thin histogram, and the thick zigged 
  curve presents an estimate of the true count behavior, $\lambda_t$.
  On the right-hand side we show the HR (hourly rate) for the 
  convenience of readers.
  }\label{fig:all}
\end{figure}

We present the count rate data of the Leonid meteor storm and its estimate 
for the true density distribution in Figure~\ref{fig:all}.
At a glance we can see that the density estimate has a fairly smooth spatial
structure, and that the violent statistical fluctuation is significantly 
suppressed by the self-organizing state space method.
The estimation uncertainty, which is simultaneously estimated by the 
recursive estimation process, is $\sim 0.1$.
The $\pm 1\mbox{-}\sigma$ uncertainty envelopes are also plotted in 
Figure~\ref{fig:all}, but it is hard to resolve on the figure.
This result shows that most of the `spikes' in the meteor count data
are merely a consequence of the Poisson fluctuation, and have no physical 
substance.
Therefore, even if a spike appears to be strong, it may be explained by
a large variance (standard deviation) of the Poisson process.

By our method, only those spikes which cannot be regarded as a mere fluctuation
are detected in the trend estimate.
The most prominent feature is the burst of meteor flux just before the first
global peak of the storm.
We can observe some other small peaks at 100, 130, 160, 210, 250, 260, and 270 min 
in Figure~\ref{fig:all}.
They clearly correspond to the peaks suggested by an analysis of \citet{shiki03}.
Interestingly, the first spike of the count at 45~min ($\sim 15^{\rm h}
25^{\rm m}$) appeared to be a sudden increase of the underlying meteor flux,
and not a spiky burst of count in the estimate.

Thus, we conclude that the true temporal trend of the dust trail which 
caused the Leonid meteor storm is globally smooth, with small and dense 
clumps associated with bursts of meteor counts.
This confirms the suggestion from a classical analysis by \citet{shiki03}, 
and provides a statistically rigorous basis on it.

\subsection{Magnitude Dependence}

\begin{figure}
  \begin{center}
  \FigureFile(80mm,80mm){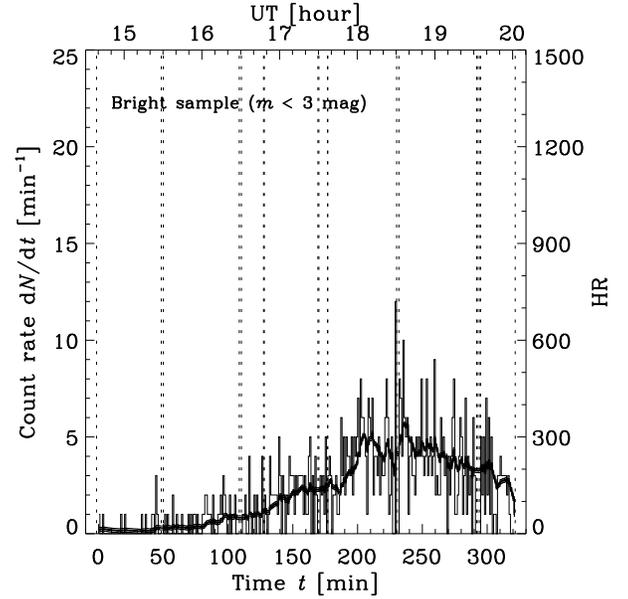}
  \end{center}
  \caption{Count rate data of the Leonid meteor storm for meteors brighter 
  than 3~mag, and its estimate for the true density distribution.
  Again, the vertical dotted lines indicate the boundaries of the data.
  The observed counts are shown by the thin histogram, and the thick zigged 
  curve presents an estimate of the true count.
  }\label{fig:bright}
\end{figure}

\begin{figure}
  \begin{center}
  \FigureFile(80mm,80mm){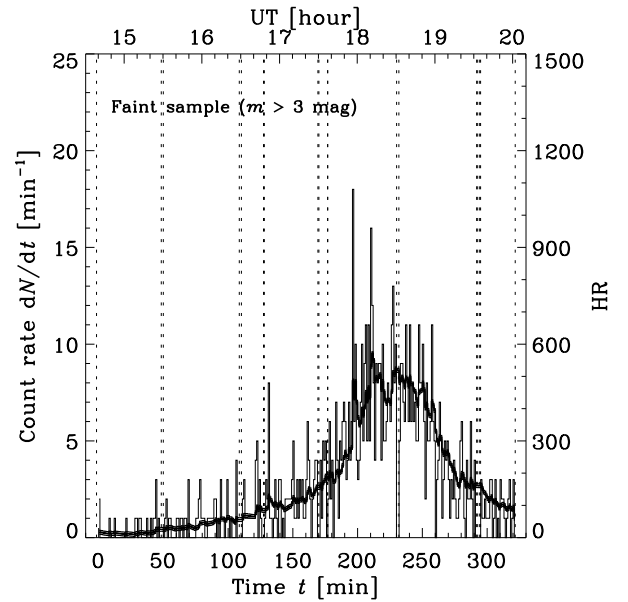}
  \end{center}
  \caption{Same as Figure~\ref{fig:bright}, except that it is for 
  meteors fainter than 3~mag.
  }\label{fig:faint}
\end{figure}

Next, we divided the data into two classes, bright and faint samples, 
and applied the self-organizing state space method to both of them,
just as we did for the whole sample.
We set the boundary of the bright and faint samples at 3~mag.
The results of the bright and faint samples are shown in 
Figures~\ref{fig:bright} and \ref{fig:faint}, respectively.

A drastic difference is found between the counts of the bright and faint 
samples.
The most striking feature is that there is no trend of bursts in bright 
sample counts at 200~min ($\sim 6^{\rm h}00^{\rm m}$~UT), whereas a prominent 
burst exists in the faint counts.
Other weaker bursts also stem from the faint count behavior.
In other words, the bright sample count is relatively smooth and its variation
is small, while the faint count temporarily varies with a very similar 
trend of the total count profile.
This clearly shows that most of the bursts have been dominated by meteors 
fainter than 3~mag.

We also find a clear excess of bright meteors to faint ones after
270~min ($19^{\rm h}20^{\rm m}$~UT).
It is an unexpected trend, because it is widely accepted that fainter
meteors are more numerous than brighter ones.
One may suspect the effect of the elevation of the radiation point, but 
this result is unchanged by the correction
because the correction has no magnitude dependence.
This suggests a bias in the distributions of larger and smaller meteorites,
which may be the origins of brighter and fainter meteors, respectively.

\citet{hughes73} reported a clear difference in the cumulative influxes of 
meteors brighter and fainter than 3~mag.
Our result may be closely related to his conclusion, but we do not go 
further here.

\begin{figure}
  \begin{center}
  \FigureFile(80mm,80mm){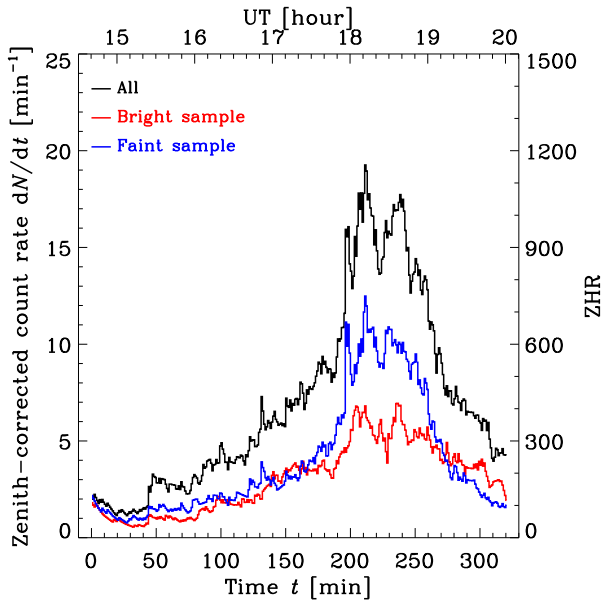}
  \end{center}
  \caption{Zenith-corrected meteor flux estimates.
  }\label{fig:zhr}
\end{figure}

So far, we have not corrected the zenith effect on the meteor flux so as to avoid
any unnecessary intricacy in statistical modeling.
If one may wish to have a zenith-corrected count rate (or equivalently, ZHR), 
we should merely make a correction of the obtained estimate here.
Figure~\ref{fig:zhr} shows the corrected meteor flux and ZHR, together
with those of bright and faint subsamples.
For the correction, we simply multiplied $1/\sin h$ ($h$ : elevation of 
the radiation point) to the flux estimates.
If we use an empirical formula of the form $1/(\sin h)^\gamma$ 
($\gamma \simeq 1.4$: see e.g., \cite{jenniskens94}), the trend around 
$15^{\rm h} \mbox{--} 17^{\rm h}$ would be more emphasized.
We note that the sum of the estimates for these subsampels perfectly
agrees with the estimates for a whole sample.
This also shows that our approach is very powerful, robust, and consistent 
for this type of analysis.
In figure~\ref{fig:zhr}, the coincidence of the burst spikes in the whole sample
and faint subsample is impressive.

\subsection{Future Prospects of the Self-Organizing State Space Model 
for Astrophysical Applications}\label{sec:prospects}

Before closing this article we would like to devote a subsection to 
some future prospects for applying the self-organizing state space model 
approach.
This approach has a very wide range of its applicability: for example, 
it can be used to estimate an extremely faint optical source variability,
and to analyze count rates of low-level photons, X-ray or cosmic ray detectors, 
which are regarded as representatives of typical count processes.

Another important aspect is its robustness against irregular sampling.
Hence, we can easily apply this method to the photometric sequence data of 
gravitationally lensed objects to measure their time delay in variability.

\citet{higuchi01} illustrated an interesting application to an estimation
of spiral density wave in Saturn's ring observed by Voyager, which is 
known to have a varying freqnency along with a radial position \citep{horn96}.
He applied the self-organizing state space model approach to the data
and showed a beautiful result.
Sunspot number data are also very popular count process 
in statistical science \citep{higuchi99}.

Thus, we expect a variety of applications of this analysis.
Now, with this approach, we do not have to worry about unrealistic 
assumptions of stationarity nor Gaussianity that are hardly expected for 
real datasets, in spite of  the fact that they are often required for the popular
classical time series analysis.
Also, we will never be annoyed by irregularly appearing observational 
gaps, sampling inhomogeneity, or heteroskedasticity that are often inherent in 
various astronomical datasets.

\section{Summary and Conclusions}\label{sec:conclusion}

The Leonids show meteor storms in a period of 33 years, and known as one of  the
most active meteor showers. 
It has recently shown a meteor stream consisting of several narrow dust trails
made by meteoroids ejected from a parent comet; hence, an analysis of the 
temporal behavior of the meteor flux is important for studying the 
structure of the trails.
However, statistical inference for the count data is not an easy task,
because of its Poisson characteristics.
We carried out a wide-field video observation of the Leonid meteor 
storm in 2001 \citep{shiki03}.

In this study, we formulated a state-of-the-art statistical analysis, 
which is called the self-organizing state space model, to infer the true 
behavior of the dust density of the trails properly from the meteor count 
data.
{}From this analysis we found that the trails have a fairly smooth spatial
structure, with small and dense clumps, which cause a temporal burst of 
meteor flux.
We also confirmed that the time behavior (trend) of the fluxes of bright 
and faint meteors are significantly different.

In summary, for the first time we obtained a reliable estimate of the true 
dust trail density profile by the self-organizing state space approach.

\bigskip
First of all, we are greatly indebted to Prof.\ Tomoyuki Higuchi, the referee, whose careful
reading and comments imporved the quality and rigor of this paper much.
T.T.T.\ also thanks Prof.\ Peter Brockwell for developing and providing 
their analysis software ITSM, which has brought a number of insights
into our pre-analysis, and Dr.\ Takako T. Ishii for detailed instruction
of the coding for the development of the self-organizing state space model.
T.T.T.\ has been financially supported by the Japan Society of the Promotion 
of Science.


\begin{thebibliography}{}
\bibitem[Akaike, Kitagawa(1999)]{akaike99}
  Akaike, H., \& Kitagawa, G.\ 1999, The Practice of Time Series Analysis
  (New York: Springer-Verlag)
\bibitem[Brockwell, Davis(1996)]{brockwell96}
  Brockwell, P. J., \& Davis, R. A.\ 1996, Introduction to Time Series and 
  Forecasting (New York: Springer-Verlag)
\bibitem[Brown et al.(2002)]{brown02}
  Brown, P., Campbell, M., Suggs, R., Cooke, W., Theijsmeijer, C., 
  Hawkes, R.\ L., Jones, J., \& Ellis, K.\ J.\ 2002, MNRAS, 335, 473
\bibitem[Cameron, Trivedi(1998)]{cameron98}
  Cameron, C.\ A., \& Trivedi, P.\ K.\ 1998, Regression Analysis of Count Data
  (New York: Cambridge University Press)
\bibitem[Durbin, Koopman(2001)]{durbin01}
  Durbin, J., \& Koopman, S.\ J.\ 2001, Time Series Analysis by State Space 
  Methods (New York: Oxford University\ Press)
\bibitem[Hughes(1973)]{hughes73}
  Hughes, D.\ W.\ 1973, MNRAS, 161, 113
\bibitem[Kalman(1960)]{kalman60}
  Kalman, R.\ E.\ 1960, J.\ Basic Eng., 82, 35
\bibitem[Kitagawa(1987)]{kitagawa87}
  Kitagawa, G.\ 1987, J.\ Amer.\ Stat.\ Assoc., 82, 1032
\bibitem[Kitagawa(1996)]{kitagawa96}
  Kitagawa, G.\ 1996, J.\ Comput.\ Graph.\ Statistics, 5, 1
\bibitem[Kitagawa(1998)]{kitagawa98}
  Kitagawa, G.\ 1998, J.\ Amer.\ Stat.\ Assoc., 93, 1203
\bibitem[Kitagawa, Gersch(1996)]{kg96}
  Kitagawa, G., \& Gersch, W.\ 1996, Smoothness Priors Analysis of 
  Time Series, Lecture Notes in Statistics, Vol.\ 116
  (New York: Springer-Verlag)
\bibitem[Kitagawa, Sato(2001)]{kitagawa01}
  Kitagawa, G., \& Sato, S.\ 2001, in Sequential Monte Carlo Methods in 
  Practice, ed.\ A.\ Douchet, N.\ de Freitas, \& N.\ Gordon
  (New York: Springer-Verlag), 177
\bibitem[Knuth(1998)]{knuth98}
  Knuth, D. E. 1998, The Art of Computer Programming, Vol. 2 Seminumerical
  Algorithms, 3rd edition (New York: Addison-Wesley)
\bibitem[Harvey(1981)]{harvey81}
  Harvey, A.\ C.\ 1981, Time Series Models 
  (Oxford: Philip Allan Publishers Ltd.)
\bibitem[Higuchi(1999)]{higuchi99}
  Higuchi, T.\ 1999, Comp.\ Stat.\ Data Analysis, 30, 281
\bibitem[Higuchi(2001)]{higuchi01}
  Higuchi, T.\ 2001, in Sequential Monte Carlo Methods in Practice, 
  ed.\ A.\ Douchet, N.\ de Freitas, \& N.\ Gordon (New York: Springer-Verlag),
  429
\bibitem[Horn et al.(1996)]{horn96}
  Horn, L.\ J., Showalter, M.\ R., \& Russell, C.\ T.\ 1996, Icarus, 124, 663
\bibitem[Jenniskens(1994)]{jenniskens94}
  Jenniskens, P.\ 1994, A\&A, 287, 990
\bibitem[Scargle(1981)]{scargle81}
  Scargle, J.\ D.\ 1981, ApJS, 45, 1
\bibitem[Shiki et al.(2003)]{shiki03}
  Shiki, S., Takeuchi, T.\ T., Miyamoto, D., Fujiwara, H., Kitazume, J., 
  \& Utsumi, Y.\ 2003, PASJ, submitted
\bibitem[Watanabe et al.(1997)]{watanabe97}
  Watanabe, J., Fukushima, H., Kinoshita, D., Sugawara, K., \& Takata, M.\ 
  1997, PASJ, 49, L35
\bibitem[Watanabe et al.(2002)]{watanabe02}
  Watanabe, J., Sekiguchi, T., Shikura, M., Naito, S., \& Abe, S.\ 2002, 
  PASJ, 54, L23
\end{thebibliography}
\end{document}